\documentclass[prl,twocolumn,showpacs,preprintnumbers,amsmath,amssymb,superscriptaddress,unsortedaddress]{revtex4}

\pdfoutput=1
\usepackage{graphicx}
\usepackage{dcolumn}
\usepackage{bm}

\begin{document}


\title{Nonadditivity of quantum and classical capacities for entanglement breaking multiple-access channels and butterfly network}

\author{Andrzej Grudka}

\affiliation{Faculty of Physics, Adam Mickiewicz University, 61-614 Pozna\'{n}, Poland}

\affiliation{National Quantum Information Centre of Gda\'{n}sk, 81-824 Sopot, Poland}

\author{Pawe{\l} Horodecki}

\affiliation{Faculty of Applied Physics and Mathematics, Technical
University of Gda\'{n}sk, 80-952 Gda\'{n}sk, Poland}

\affiliation{National Quantum Information Centre of Gda\'{n}sk, 81-824 Sopot, Poland}

\date{\today}

\begin{abstract}
We analyze quantum network primitives which are entanglement-breaking. 
We show superadditivity of quantum and classical capacity regions for 
quantum multiple-access channel and quantum butterfly network.
Since the effects are especially visible at high noise they 
suggest that quantum information effects may be particularly helpful 
in the case of the networks with occasional high noise rates. 
To our knowledge the present effects provide the first qualitative borderline 
between superadditivities of bipartite and multipartite systems.
\end{abstract}

\maketitle
{\it Introduction.} 
Fundamental discoveries of quantum cryptography  
without \cite{BB84} and with entanglement \cite{Ekert},
quantum dense coding \cite{dense} and quantum teleportation \cite{tel} 
constitute cornerstones of the domain called {\it quantum channel theory}
\cite{BennettShor,Bennett-capacities}. Very important,  purely quantum phenomena 
are superadditivities of capacities in multipartite variants of quantum 
capacity $Q$ with classical side-channel \cite{Duer} (cf. \cite{superactivation}). 
One of the newly observed effects was nonadditivity of classical capacity $C$ of multiple-access channel with no side resource \cite{Czekaj}
(see \cite{Czekaj1} for continuous variables analog).
Recently, a fundamental, most striking superadditivity in bipartite scenario for quantum capacity $Q$ with 
no side resources was discovered \cite{Smith} and followed by an announcement 
of another surprising phenomenon of breaking  additivity of secret key capacity $K$ \cite{Winter} which can be refined to extreme 
cases \cite{Smith-Smolin2} (cf. \cite{Smith-Smolin1}). 
Very challenging open problems is additivity of classical capacity $C$ 
in bipartite scenario. The conjecture of 
additivity of so called Holevo capacity $\chi(\Lambda)$ has been disproved recently 
in an impressive way \cite{Hastings} where superadditivity for two 
channels was proven. The problem of additivity of the capacity $C(\Lambda)$
is still open since the latter is an asymptotic quantity.
During the research on that fascinating issue it has
been shown in particular that bipartite channels which are entanglement-breaking
\cite{EntanglementBreaking} (i.e., channels which cannot create entanglement between sender and receiver)
cannot contribute to such superaditivity phenomena \cite{EntanglementBreaking, oneway, Shor1}.
 
In the present paper we address the question whether superadditivity of capacity of entanglement breaking channels
is valid in multipartite scenarios. We find, quite surprisingly, that   
it is not true: both $Q$ and $C$ (i.e., quantum and classical capacities without side resources) in the case of two-access entanglement breaking channels 
may exhibit superadditivity when supplied with a highly entangling channel. We show also strong nonadditivity of capacity in the quantum butterfly network  \cite{q-butterfly}.
{\it None} of the present effects can have an analog in bipartite scenarios.
In this way, our result provides for the first time the superadditivity effects sharply 
discriminating between bipartite scheme and that with more than two users.

For classical capacities our quantum networks violate a special rule which is valid 
for all discrete classical networks and follows 
immediately from the additivity theorem provided in \cite{Czekaj}: {\it in any classical
multiple-access network primitive it is impossible to improve the transfer rate 
of one sender by adding resources to another sender.}
Here we shall call it the locality rule (LR) of data transfer.

{\it Multiple-access entanglement breaking channels and superaddivitity.} 
Let us present a pair of channels for which one has superadditivity of quantum capacity. The first channel is presented in Fig. 1.  Alice and Bob have $d$ dimensional inputs, while Charlie has $d$ dimensional output. The channel performs the Bell measurement on two qudits and sends a result of the measurement to Charlie. Formally our channel can be written as a completely positive trace preserving linear map
\begin{eqnarray}
& \Lambda(\varrho_{AB})=\nonumber\\
& =\sum_i\text{Tr}_{AB}(|\Psi^i\rangle \langle \Psi^i|_{AB}\varrho_{AB}|\Psi^i\rangle \langle \Psi^i|_{AB})|\Psi^i\rangle \langle \Psi^i|_C,
\end{eqnarray}
where $|\Psi^i\rangle_{AB}$ are $d^2$ orthogonal Bell states. Because $|\Psi^i\rangle \langle \Psi^i|_{AB}$ are Kraus operators of rank one, the channel is entanglement breaking. Hence, the quantum capacity region of this channel is given by $R_A = 0$ and $R_B = 0$.
The second channel is the identity qudit channel from Bob to Charlie. Its quantum capacity region is given by $R_A = 0$ and $R_B \leq \log d$.

\begin{figure}
\includegraphics [width=6truecm]{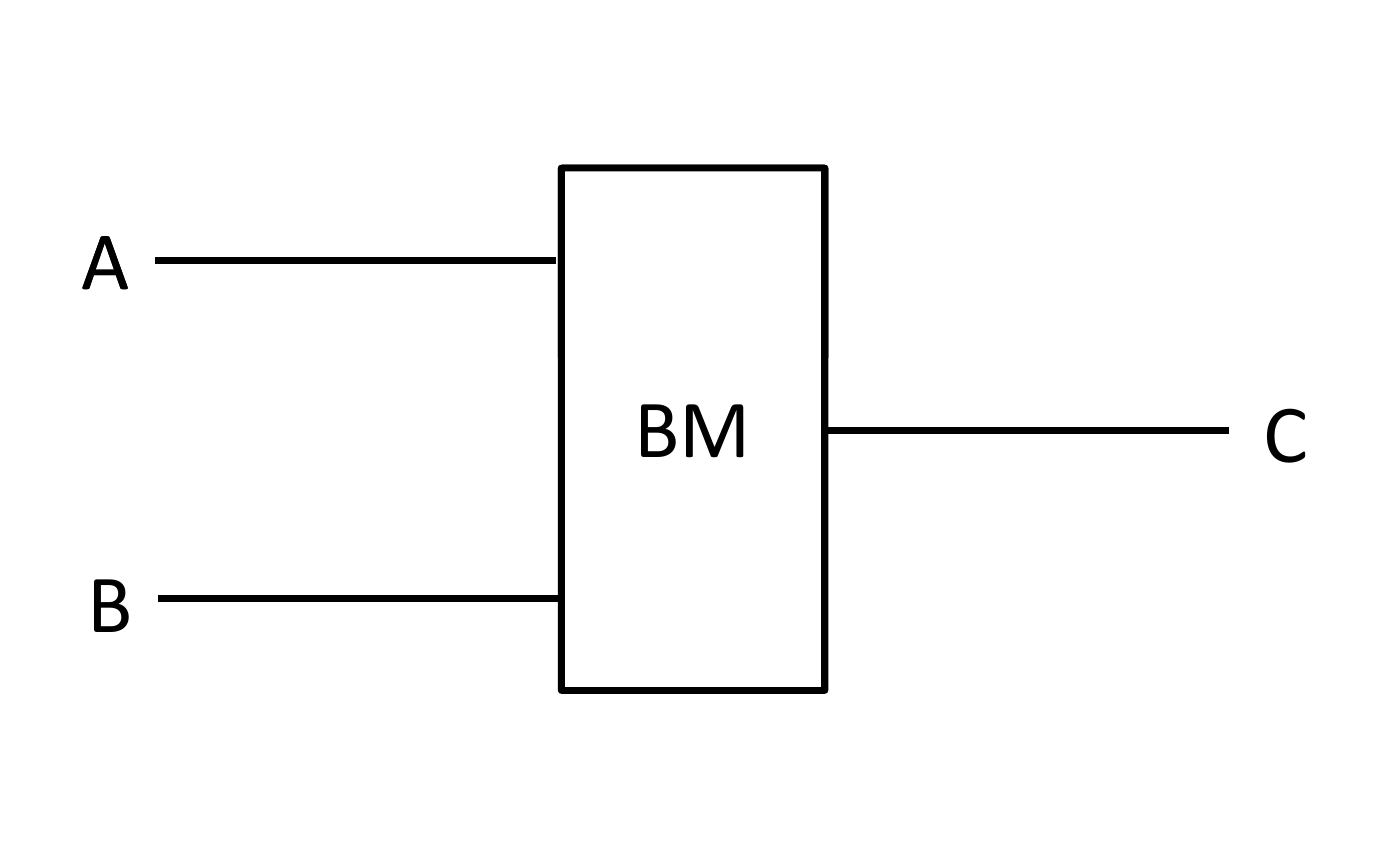}
\caption{\label{fig:1} Entanglement breaking multiple-access channel. $BM$ stands for Bell measurement.}
\end{figure}

We now find quantum capacity region of the tensor product of these two channels. Let Bob send half of the maximally entangled pair of qudits through the first channel and the other half through the second channel and let Alice send a qudit through the first channel. Because the first channel measures a qudit sent by Alice and a qudit from the maximally entangled state in the Bell basis and sends a result of the measurement to the receiver, it effectively teleports a qudit sent by Alice to the output of the second channel. Hence, the rate pair $(R_A, R_B)=(\log d,0)$ can be achieved. On the other hand, $R_A+R_B$ cannot be greater than $\log d$ because the first channel performs the complete von Neumann measurement on two qudits. In consequence, the  quantum capacity region of the tensor product of these two channels is given by 
\begin{eqnarray}
& R_A+R_B \leq \log d.
\end{eqnarray}

Our channel is an entanglement breaking channel in contrast to the channel considered in Ref. \cite{Czekaj}, which shows nonadditivity of classical capacity regions. One may wonder if it is possible to show nonadditivity of classical capacity regions for entanglement breaking channel and some other channel. Below we demonstrate such a pair of channels. The first channel is presented in Fig. 2. Alice and Bob have $d^2$ and $d$ dimensional inputs, respectively, while Charlie has $d$ dimensional output.  The channel transmits a qudit from Bob to Charlie. Depending on the state of Alice's qudit, the state of Bob's qudit is transformed by one of $d^2$ unitary operations used in dense coding protocol. After this transformation, Bob's qudit is sent through the depolarizing channel
\begin{eqnarray}
D_x(\varrho)=(1-x)\rho+x\frac{I}{d},
\end{eqnarray}
while Alice's qudit is discarded.
For $x \geq \frac{d}{d+1}$, the depolarizing channel, and hence also our channel, is entanglement breaking.
The classical capacity region of this channel is given by $R_A+R_B \leq C$.
$C$ is Holevo capacity of the depolarizing channel $D_x$ and is given by formula
\begin{eqnarray}
C=\log d -H_d (1-x\frac{d-1}{d}),
\end{eqnarray}
where $H_d(x)=-x\log x -(1-x) \log\frac{1-x}{d-1}$.
The second channel is the identity qudit channel from Bob to Charlie. Its classical capacity region is given by $R_A = 0$ and $R_B \leq \log d$.

\begin{figure}
\includegraphics [width=6truecm]{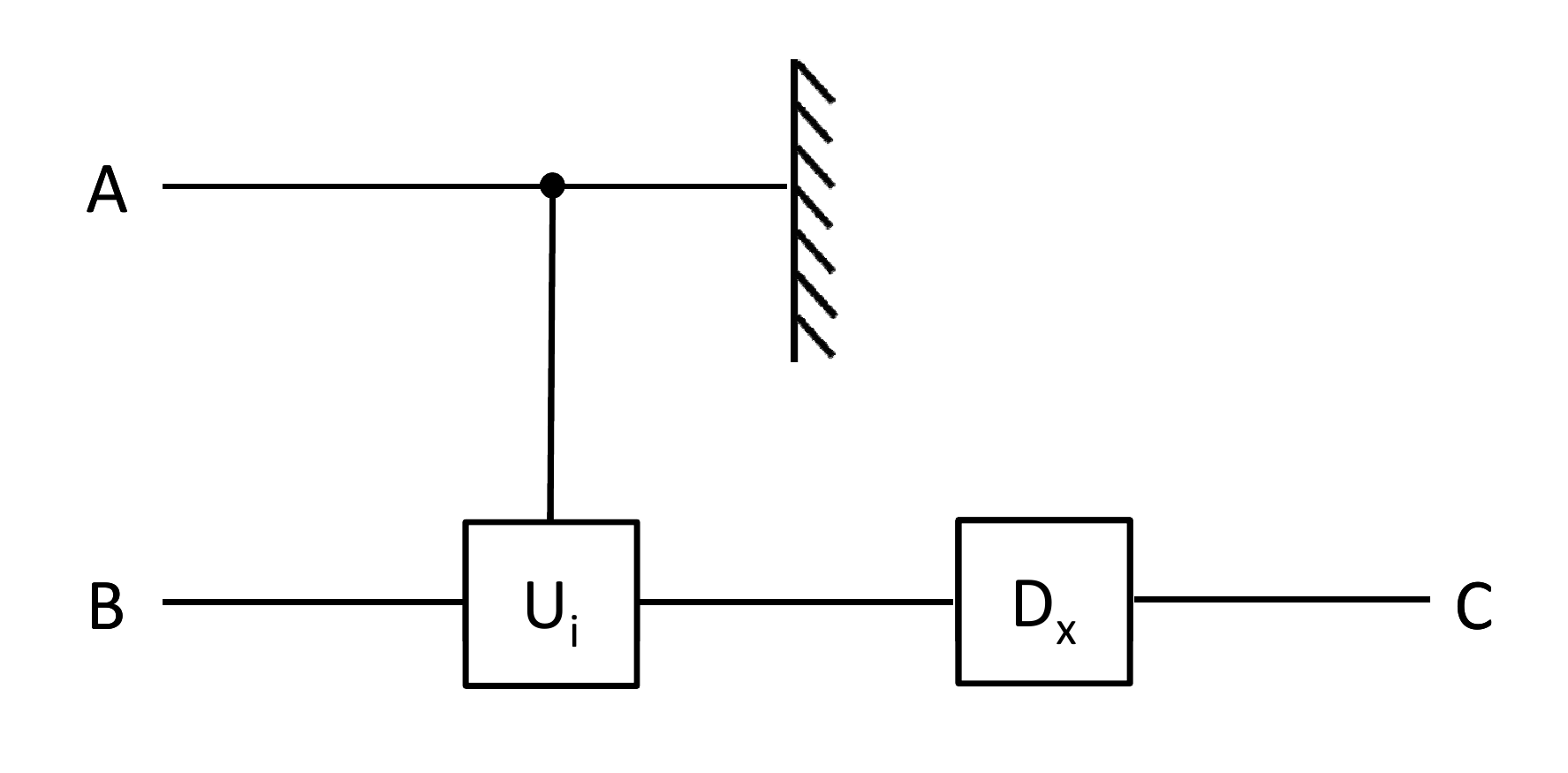}
\caption{\label{fig:2} Entanglement breaking multiple-access channel. $U_i$ stands for controlled unitary operation, $D_x$ stands for depolarizing channel.}
\end{figure}

We now turn our attention to classical capacity region of the tensor product of these two channels. When  Bob sends half of the maximally entangled pair of qudits through the first channel and the other half through the second channel, then Alice can transform the maximally entangled state to one of $d^2$ orthogonal states by inputing to the first channel one of $d^2$ orthogonal states. Because the first qudit from the maximally entangled state is sent through the depolarizing channel and the second qudit is sent through the identity channel, the parties can achieve in this way the rate pair $(R_A, R_B)=(C_E,0)$. $C_E$ is entanglement assisted classical capacity of the depolarizing channel \cite{EntanglementAssisted} and is given by formula
\begin{eqnarray}
C_E= 2 \log d -H_{d^2}(1-x\frac{d^2-1}{d^2}).
\end{eqnarray}
Alice cannot send more than $C_E$ bits of information as she does not control the input to the second channel and hence the entanglement assisted classical capacity of the depolarizing channel is the maximal capacity which can be achieved. On the other hand,  $R_A + R_B \leq \log d+C$ because it cannot be greater than Holevo capacity of the tensor product of the depolarizing channel and the identity qudit channel. Hence, two extreme points of classical capacity region of the tensor product of these two channels are given by
\begin{eqnarray}
& (R_A, R_B) =(C_E, 0), \nonumber\\
& (R_A, R_B) =  (0,C + \log d).
\end{eqnarray}
These extreme points prove nonadditivity of capacity regions. If $x \rightarrow 1$ then $C_E/C \rightarrow d+1$ and we can have arbitrarily large superadditivity of the capacity regions.

{\it Noisy extensions.}
It is worth noting that one can consider two natural modifications 
of the channel which demonstrate nonadditivity of quantum capacity.
(i) The first one is a mixture of the Bell measurement which happens with probability $1-q$ and 
classical uniform noise which happens with probability $q$.
Together with the identity qudit channel from Bob to Charlie, this channel can simulate the quantum depolarizing channel $D_{q}$ from Alice to Charlie.
In fact, with probability $1-q$ Charlie can completely recover a quantum message while with probability $q$ he is left with the completely 
random noise coming from part of the singlet state (apart from completely useless classical uniform noise).
Hence, in this case one can achieve $R_A = Q(D_q)$. 
(ii) Suppose that instead of the just descibed channel, we have a mixture (with the same probabilities)
 of the Bell measurement and the identity channel. The channel also returns a flag marking which of the two events happened. If this channel is supported by the identity qudit channel from Bob to Charlie then one can achieve  $R_A=\log d$.

{\it General networks: amplifying swapping transfer and quantum version of the butterfly network.}
Consider the channel $\Phi$ provided in Fig. 3. Each 
sender has $d^2$-dimensional classical input and $d$-dimensional quantum one.
Since here we deal with quantum channels which have more than one sender, we may also include the {\it  common information } rate \cite{CoverThomas}, i.e., the rate of 
the same information that is faithfully transfered to both receivers $\tilde{A}$, $\tilde{B}$.
We denote the common information rate by $R_{X}^{(o)}$, where $X=\{A,B\}$ stands for 
the single sender's system or, more generally, the sender's site which may contain many systems at the local sender disposal. 
The total rate vector is denoted by  ${\bf  R}=(R_{A\tilde{A}},R_{A\tilde{B}},R_{B\tilde{B}},R_{B\tilde{B}}
,R_{A}^{(o)},R_{B}^{(o)})$.
We must stress here that this description is more detailed than the one usually used (cf. \cite{q-butterfly}).
In fact, one often analyzes only rates $R_{A\tilde{B}}$, $R_{B\tilde{A}}$ 
for the fixed values of $R_{A\tilde{A}}$, $R_{B\tilde{B}}$ which are assumed to contain also 
common information which is not counted separately. We keep here all rates 
since it is more natural taking into account the structure of the channel we consider.
\begin{figure}
\includegraphics [width=6truecm]{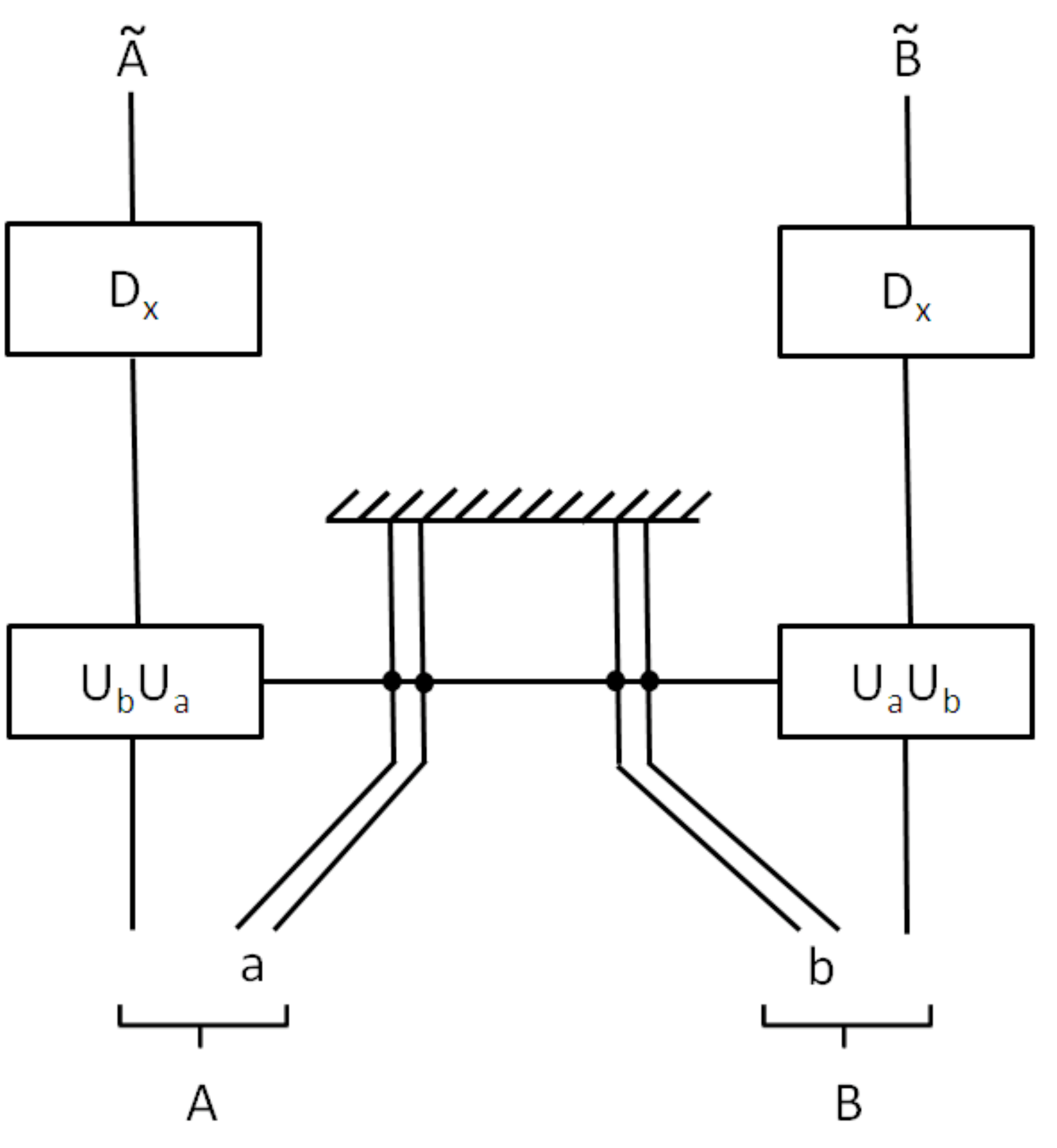}
\caption{\label{fig:3} Entanglement breaking quantum butterfly network. $U_a$ and $U_b$ stand for controlled unitary operations. $D_x$ stands for depolarizing channel.}
\end{figure}
From the fact that just before both outputs of the channel we have depolarizing channels 
$D_{x}$, it follows that the total capacity region of the channel 
is contained in the set ${\cal S}$ satisfying the following conditions:
\begin{eqnarray}
&R_{A\tilde{A}} + R_{B\tilde{A}} + R_{A}^{(o)} \leq C,\nonumber \\  
&R_{A\tilde{B}} + R_{B\tilde{B}} + R_{B}^{(o)} \leq C.
\label{initial-region}
\end{eqnarray}
Thus, we have in  short $C(\Phi)\subset \cal{S}$.
Suppose now that we assist the channel with product of two identity channels
$\Theta_{A'B'\rightarrow \tilde{A}' \tilde{B}'}=I_{A'\rightarrow \tilde{A}'} \otimes I_{B'\rightarrow \tilde{B}'}$.
This channel has clearly the transmission rate region $C(\Theta)$:
\begin{eqnarray}
&R_{A'\tilde{A}'}\leq \log d, \nonumber \\
&R_{B'\tilde{B}'}\leq \log d.
\end{eqnarray}

\begin{figure}
\includegraphics [width=6truecm]{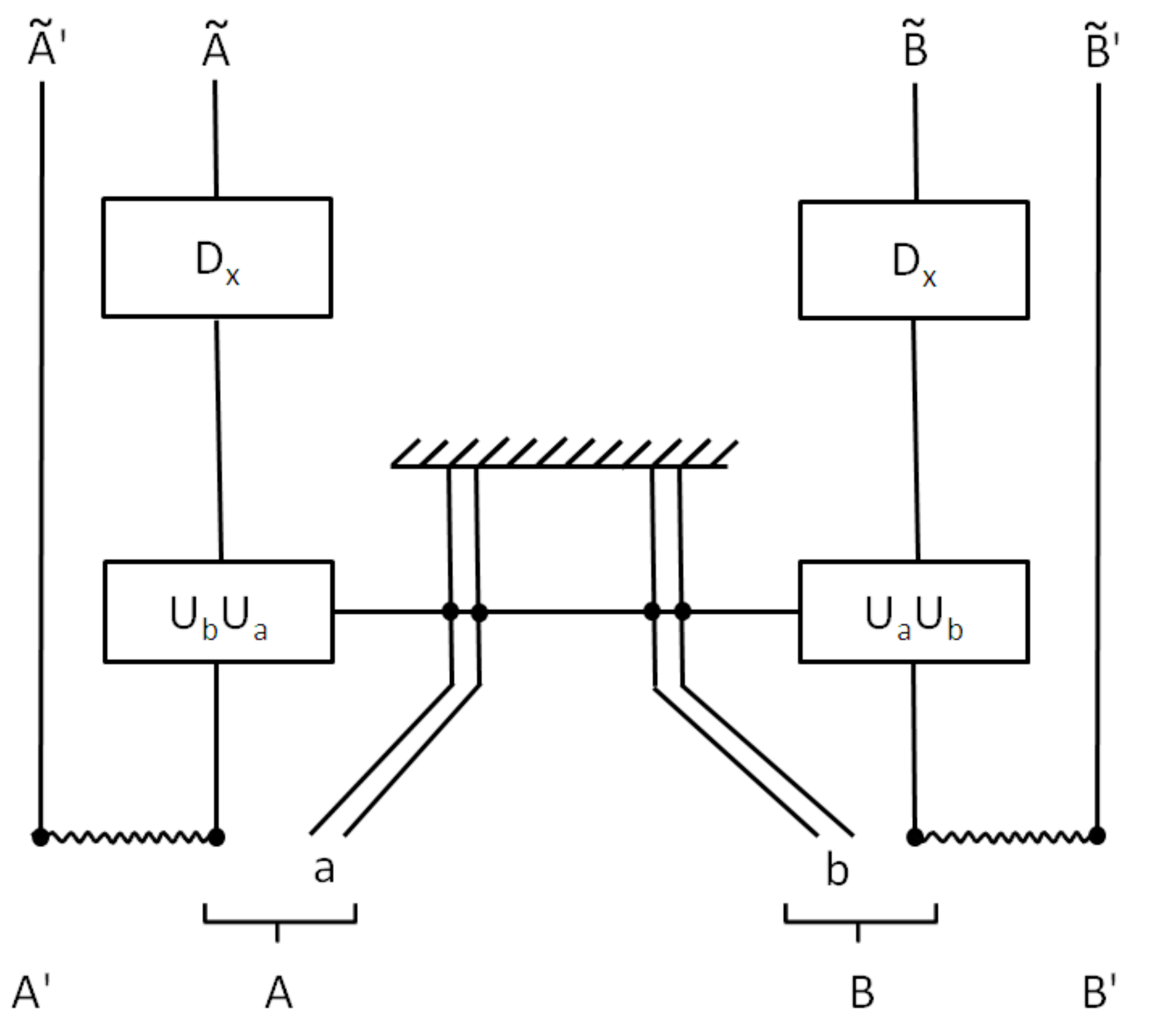}
\caption{\label{fig:4} Entanglement breaking quantum butterfly network assisted by two identity channels.}
\end{figure}

Consider the special strategy achieving particularily interesting transmission rates
for the butterfly network from Fig. 3 assisted by two identity channels 
(see Fig. 4 for the assistance scheme).
Any message $a$ by Alice and $b$ by Bob can be sent down their classical input of 
the channel and at the same time can be encoded by 
$( ( U^{a}_{A}  )^{\dagger}\otimes I_{A'})|\Psi^{+}\rangle_{AA'} $ and $(  ( U^{b}_{B} )^{\dagger} \otimes I_{B'})|\Psi^{+}\rangle_{BB'} $, where $|\Psi^{+}\rangle_{XX'}$ is $d \otimes d$ 
maximally entangled state and $U_X^x$ is one of $d^2$ unitary operations which are used in dense coding protocol. 
The channel will effectively send  the first 
half of the state $(U^{b}_{A} \otimes I_{A'})|\Psi^{+}\rangle_{AA'}$ through the depolarizing channel and the second  half through the identity channel to Alice's receiver's side and at the same time it will effectively send the first half of the state $(U^{a}_{B} \otimes I_{B'})|\Psi^{+}\rangle_{BB'}$ through the depolarizing channel and the second half through the identity channel to Bob's receiver's site.
However, each of the states is just the same as if it was coming out of bipartite entanglement 
assisted quantum depolarizing channel as in the previous paragraph.
Hence, both senders achieve now the cross-transfer rates (here the 
subscripts denote sides and not the systems which must have been marked by additional $\tilde{X}$ 
notations):
\begin{eqnarray}
&& R_{A\tilde{B}}=C_{E}, \nonumber \\
&& R_{B\tilde{A}}=C_{E}, 
\label{x-region}
\end{eqnarray} 
where $C_{E}>C$. The other rates in vector ${\bf R}$ are equal to zero in the case of these states.

To compare the effect with the classical case, we should prove that 
it is impossible in a classical network. To show this, 
consider the part of the network with one of the receivers 
traced out, for example tracing out Bob's receiver's
parts $\tilde{B}$, $\tilde{B}'$.
Since the input local messages are independent, a classical analog of such 
a remaining network primitive (i.e. the one with two senders and one receiver) must obey 
our locality rule (LR). This says immediately that all what
the classical network may offer in bits transmitted from $B$ to $\tilde{A}$ in this 
case, is $C$ in (\ref{x-region}) (instead of $C_{E}$)
which is the original bound (\ref{initial-region}).
To see it more clearly, let us notice that the additional noiseless $d$-ary forward 
channel from $A'$ to $\tilde{A}'$ cannot     
improve the transfer rate $R_{B\tilde{A}}$ (according to the locality rule applied to this $2$-access 
channel with two senders $A$, $B$, and one receiver $\tilde{A}$), so the latter must remain equal to $C$ as if the 
new connection $A' \rightarrow \tilde{A}'$ did not exist.
The above remark is completely independent of possible internal machinery 
of the $2$-access channel considered as long as it is classical. 
In particular, it is obeyed by the original XOR gate. This clearly proves
that superadditivity of this type cannot happen in classical networks.

{\it Conclusions.}
Superadditivities of all kinds found so far in quantum scenarios (irrespectively of 
whether they were bipartite or multipartite) required that both channels
may create entanglement. For quantum capacities
this rule is well understood in the case of bipartite scenario. 
On the one hand, entanglement breaking channel can be simulated with the help of forward classical communication \cite{EntanglementBreaking}. On the other hand, forward classical communication  cannot increase quantum capacity of the channel \cite{oneway}.
For classical capacities of bipartite channels the rule was proven independently in 
\cite{Shor1}, where it was shown that Holevo capacity is additive on a tensor product of two channels, when one of the channels is entanglement breaking.
It could be expected that the rule could be generalized to multipartite networks. Here we have shown that this is {\it not} the case.
We have considered two types of primitives for quantum networks:
$2$-access channels, i.e. one with two senders and one receiver and the butterfly network. We have proven that even if one channel or network is entanglement breaking,
the superadditivity effect may still hold for both classical and quantum capacities if other 
chanels have their transmission rates good enough (the identity channels may be perturbed by small 
noise and still our results hold by simple continuity arguments).
 Usually one looks for the effects that discriminate between different types 
 of communication resources. For instance, multipartite entanglement is different from bipartite entanglement 
 since there are nonequivalent types of multipartite entanglement (GHZ and W states). 
We may ask about the qualitative differences between bipartite and 
 multipartite communication. So far it seemed that all superadditivity effects 
 found in the multipartite case had their, much harder to find, but of similar type, 
 analogs in bipartite scenario.
The present superadditivity effects 
 for entanglement-breaking channels are the first ones that sharply discriminate 
 between bipartite and multipartite scenarios, i.e., they cannot happen 
 in bipartite scenarios. Finally, we note that the size of the amplification 
 at high noise rates makes it interesting for applications in occasionally 
 very noisy communication systems.

\begin{acknowledgments}
We thank Joanna Mod{\l}awska for preparation of figures. We thank Fernando Brand\~ao and Graeme Smith for valuable comments on the manuscript. This work was supported by IP SCALA project and partially by  LFPPI network.  A.G. was partially supported by Ministry of Science and Higher
Education Grant No. N N206 2701 33. 
\end{acknowledgments}


\end{document}